\documentclass[aps,prb,twocolumn]{revtex4-1} 

\usepackage[dvips]{graphicx}
\usepackage{color}
\usepackage{psfrag}

\usepackage{amsthm}
\usepackage{amsmath}
\usepackage{amsfonts}
\usepackage{amssymb}

\DeclareMathOperator{\Li}{Li}

\renewcommand{\vec}[1]{\mathbf{#1}}

\def\vp{\vec{p}}
\def\vA{\vec{A}}

\def\tt{\tilde{t}}

\def\vE{\vec{E}}

\def\vsigma{\boldsymbol{\sigma}}

\def\jpc{j_{\vp}^{\text{c}}}
\def\jpp{j_{\vp}^{\text{p}}}
\newcommand{\jint}[2]{j_{#2}^{\text{#1}}}

\begin{document}
\title{Nonequilibrium transport and statistics of Schwinger pair production in Weyl semimetals}

\author{Szabolcs Vajna}
\affiliation{Department of Physics and BME-MTA Exotic  Quantum  Phases Research Group, Budapest University of Technology and
  Economics, 1521 Budapest, Hungary}
\author{Bal\'azs D\'ora}
\affiliation{Department of Physics and BME-MTA Exotic  Quantum  Phases Research Group, Budapest University of Technology and
  Economics, 1521 Budapest, Hungary}
\author{R. Moessner}
\affiliation{ Max-Planck-Institut f\"ur Physik komplexer Systeme, N\"othnitzer Str. 38, 01187 Dresden, Germany}

\begin{abstract}
The non-equilibrium dynamics beyond linear response of Weyl semimetals is studied after a sudden switching on of a DC electric field. 
The resulting current is a nonmonotonic function of time, with an initial quick increase of polarization current followed by a 
power-law decay. Particle-hole creation \`a la Schwinger dominates for long times when the conduction current 
takes over the leading role, with the total current increasing again. 
The conductivity estimated from a dynamical calculation within a Drude picture agrees with the one obtained from Kubo's formula.
 The full distribution function of electron-hole pairs changes from Poissonian for short perturbations 
to a Gaussian in the long perturbation (Landau-Zener) regime. 
The vacuum persistence probability of high energy physics manifests itself in a finite probability of no pair creation and no induced current at all times.
\end{abstract}
\date{\today}
\maketitle{}  
\bibliographystyle{apsrev}  

\section{Introduction}
Condensed matter systems, e.g. graphene, 3D topological insulators and Weyl semimetals, provide unique opportunity to examine fascinating QED effects, like Klein 
tunneling, Zitterbewegung, chiral anomaly or Schwinger pair production, most of which barely accessible to experiment otherwise. In addition to this ``fundamental'' appeal, these phenomena play a crucial role in transport properties of these systems.

Weyl semimetals (WSMs) are 3D materials, which similarly to the 2D Dirac electrons in  graphene, are characterized by linearly dispersing low energy excitations around 
some points in the Brillouin zone \cite{WanPRB2011,BurkovPRL2011,HosurPRL2012,BurkovJPCM2015}. These Weyl points are intersections of nondegenerate 
bands, and are stable against perturbations according to their topological nature. The low-energy physics of these materials mimic the Weyl fermions 
well-known from high energy physics, giving the name WSM. 

Similarly to clean graphene, when the Fermi energy in WSMs is near the Weyl point, there are no charge carriers available for transport at zero 
temperature, since the density of states vanishes as $\sim \epsilon^2$ close to the Weyl point. 
However, in an applied electric field, particle-hole pairs created by the Schwinger mechanism \cite{SchwingerPR1951} contribute 
to transport. 

The non-equilibrium state that evolves after turning on an electric electric field can be characterized by the statistics of the excitations, 
and by the induced current. As pair creation is described by the Landau-Zener (LZ) formula in the strong electric field regime, 
it is intrinsically related to the Kibble-Zurek 
mechanism  \cite{PolkovnikovPRB2005,ZurekPRL2005,DamskiPRA2006}, which describes the universal scaling of defect generation in driven systems near a critical point. 
Alas, KZ scaling gives only the mean number of excitations, and thus does not fully characterize the non-equilibrium state. 

Such a characterization, however, is  possible through all the higher moments or cumulants, as these contain all information about
non-local correlations of arbitrary order  and entanglement.
This is practically equivalent to determining the full distribution function of the quantity of interest.
Therefore, the full distribution function of the number of electron-hole pairs is also of interest, yielding additional information about the physics
of Schwinger pair production. Condensed matter physics and cold atomic systems thus provide a unique way to experimentally detect
such quantities \cite{TarruellNat2012,HeWeyl2015}, beyond the current capabilities of high energy physics.
 These ideas  also relate to the discipline of full counting 
statistics \cite{nazarov2003quantum, EspositoRMP2009}, were outstanding experiments measure whole distribution 
functions\cite{MaisiPRL2014,MalossiPRL2014}, and cumulants up to the 15th order e.g. in Ref.~\cite{FlindtPNAS2009}.

Our results on the time evolution of the current and statistics of electron-hole pairs in is summarized in TABLE~\ref{tab:summary}. 
The time domain is split into three distinct regions with different behaviour, which we call classical (ultrashort), Kubo (short), and Landau-Zener regime (long perturbations).
\begin{table}[h!]
 \centering
\begin{tabular}{| c | c | c | c |}
 \hline
Time domain &  Classical & Kubo   & Landau-Zener  \\ 
 & $t\ll \frac{\hbar}{v_F \Lambda}$ & $\frac{\hbar}{v_F \Lambda} \ll t \ll \sqrt{\frac{\hbar}{v_F e E}}$ & $\sqrt{\frac{\hbar}{v_F e E}} \ll t$  \\ \hline
\# pairs ($n$) &  $\sim E^2 t^2 \Lambda$ & $\sim E^2 t$ & $\sim E^2 t$ \\ \hline
Statistics &  Poissonian & Poissonian & Gaussian-like\\ \hline
Current ($j$) &  $\sim E t \Lambda^2$ & $\sim {E}/{t}$ & $\sim E^2 t$  \\ \hline
\end{tabular} 
\caption{The electric field and time dependence of the total number of excitations or pairs created ($n$) and its statistics, together with the electric  current ($j$) is shown. 
$\Lambda$ is the momentum cutoff, $E$ is the electric field.}
\label{tab:summary}
\end{table} 

The time evolution of the current also allows us to conjecture qualitatively the behaviour of the steady state current-voltage characteristics.
For small voltages, the dynamical calculation combined with Drude theory reproduces the results of Kubo formula calculations, i.e. the current is proportional to the electric 
field. However, Ohm's law breaks down for larger voltages and the current-electric field dependence becomes non-linear. This critical electric field as well as the non-linear 
current-voltage relation are important for possible transport experiments in WSMs. 

The paper is structured as follows. First, we 
introduce the model and its solution in section~\ref{sec:hamsol}. Then we discuss the evolution of the current 
and its implications for the steady state conductivity in sections \ref{sec:currevol} and \ref{sec:Drude}. The statistics 
of pair creation is studied in section~\ref{sec:statistics}, and it is compared with a complementary measure, the vacuum persistence probability, in section~\ref{sec:vpp}.

\section{Electric field switch-on in a Weyl semimetal}\label{sec:hamsol}

We consider noninteracting Weyl fermions near a single Weyl point. A homogeneous electric field is switched on at $t=0$, which is described by a time dependent vector potential $\vA(t)=(e E t \Theta(t),0,0)$. 
The time evolution of a given mode $\vp = (p_x, p_y, p_z)$ is governed by the Hamiltonian
\begin{align} \label{eq:HamLZ}
H=v_f (\vp-e\vA(t))\cdot \vsigma \,,
\end{align} 
where $\vsigma$ denotes the vector of Pauli matrices and $v_f$ is the Fermi velocity. 
The spectrum consists of two bands as $\pm v_f\sqrt{p_x^2+p_{\perp}^2}$, with  $p_{\perp}=\sqrt{p_y^2+p_z^2}$ the perpendicular momentum. 
Initially ($t<0$), the system is 
assumed to be in the $T=0$ vacuum state, with all modes with negative single particle energy filled and positive energy modes empty.
 This effective Weyl theory is valid at low energies compared to a high energy cutoff $v_F \Lambda$ introduced for integrals over momentum space whenever necessary.

At $t=0$, the electric field is switched on, and the time dependent Schr\"odinger 
equation can be solved analytically using parabolic cylinder functions \cite{GavrilovPRD96,VitanovPRA96,TanjiAnnPhys2009}. 
The instantaneous eigenenergies form two bands as  $\pm\epsilon(p)$ with $\epsilon(p)=v_f\sqrt{(p_x-e A(t))^2+p_\perp^2}$.

The current contribution from a given mode 
$\vp$ is determined by the mode excitation probability $n_{\vp}(t)$, which gives the number of electrons created in the upper band due to the electric field and also the holes in the lower band, with 
$n_{\vp}(t=0)=0$.
The current consists of a conduction (intraband) and a polarization (interband) part as $\langle j_x\rangle_{\vp}(t)=\jpc(t)+\jpp(t)$ \cite{DoraPRB2010,DoraPRB2011}
\begin{align} \label{eq:jpc}
\jpc(t)&=-e v_F \left[\frac{v_F (p_x-eEt)}{\epsilon_{\vp}(t)} (2n_{\vp}(t)-1) \right] \\ \label{eq:jpp}
\jpp(t)&= e v_F \frac{2\epsilon_{\vp}(t)}{v_F e E} \partial_t n_{\vp}(t) 
\end{align} 
The total contribution of a Weyl node is obtained after momentum integration.
In Eq. \eqref{eq:jpc}, the  $n_{\vp}$
 independent background  is discarded, as an empty or fully occupied band does not carry current\cite{ashcroft,DoraPRB2010}. 
In our non-interacting model, the total current, excitation numbers and higher cumulants are additive, i.e. given by the sum over the Weyl nodes.

The vanishing gap is a signature of the ``criticality'' of the WSM phase. As such, it exhibits scaling properties, which allow us to deduce important properties of the system without explicitly solving the Schr\"odinger equation. The excitation probability of the modes satisfies a scaling relation (in units of $\hbar,v_F,e=1$), 
\begin{align} \label{eq:np_scale}
n_{\vp}^E(t)=n_{b \vp}^{b^2 E}(b^{-1}t) \,,
\end{align} 
which follows from the time dependent Schr\"odinger equation, and holds for any choice of the dimensionless scaling parameter $b$. 
The invariants of the scaling transformation yield the natural dimensionless combinations which determine the physics e.g. $\frac{p}{eEt}$, $\sqrt{\frac{v_F}{\hbar eE}}p$,  $\tt=\sqrt{\frac{v_F e E}{\hbar}}t$, etc. The dimensionless time $\tt=\frac{t}{t_E}$ uniquely classifies the excitation probability as a function of $\vp$, where $t_E=\sqrt{{\hbar}/{v_F e E}}$ is the time scale related to the electric field. Time reversal considerations also give constraint on the excitation probabilities \cite{TanjiAnnPhys2009}
\begin{align}
n_{\vp}(t)=n_{e{\vE}t-\vp}(t) \,,
\end{align}
which means that the excitation probability is symmetric with respect to $p_x=\frac{1}{2}e E t$. Accordingly, in Eq.~\eqref{eq:npBalazs}, and everywhere where spherical coordinates are used, the momentum is measured from $(eEt/2,0,0)$. That is, $p=\sqrt{(p_x-eEt/2)^2+p_{\perp}^2}$.

\begin{figure}[h!]
\centering
\includegraphics[width=8.4cm]{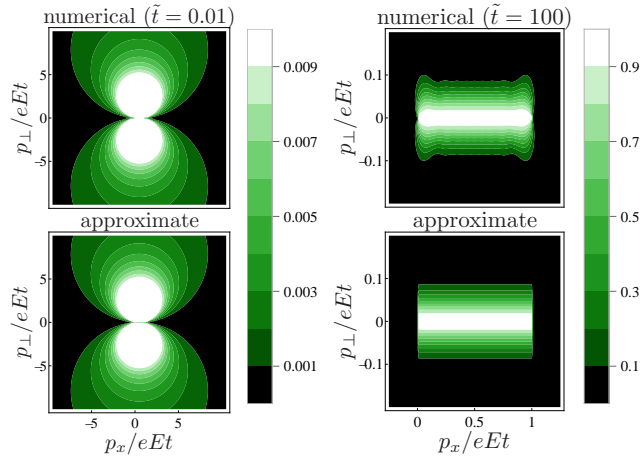}
\caption{Illustration of the excitation probabilities for short (left) and long perturbations (right). The excitation map has a ``dipolar'' character for short times, and the approximate formula \eqref{eq:npBalazs} is nearly indistinguishable from the numerical solution for $p\gg eEt$. The excitation map is cylindrical for long times. An (asymptotically irrelevant) increased number of excitations at $p_x=0$ and $p_x=e E t$ is not captured in the approximation \eqref{eq:npLZ}.}
\label{fig:np_plane}
\end{figure}

The excitation probability as a function of $p$ is qualitatively different in the $\tt\gg 1$ and 
$\tt\ll 1$ cases (Fig.~\ref{fig:np_plane}). A perturbative solution valid for $\tt\ll 1$ is~\cite{DoraPRB2010}
\begin{equation} \label{eq:npBalazs}
n_{\vp} = \frac{(e E \hbar p_{\perp})^2}{4 v_F^2 p^6}\sin^2\left(\frac{v_F p t}{\hbar}\right) \,.
\end{equation}
 This gives a good approximation for the excitation number for $p \gg e E t$. At short times high energy 
states may become excited, which is reflected in the power law decay of excitations as a function of momentum ($\sim p^{-2}$ for $p\ll \hbar/v_F t$). 

If the perturbation is long, the probability of exciting a given mode is well approximated by the LZ solution \cite{HallinPRD1995}.
\begin{align} \label{eq:npLZ}
n_{\vp}&=\Theta(p_x)\Theta(e E t-p_x) \exp\left(-\dfrac{\pi v_f p_{\perp}^2}{\hbar e E}\right).
\end{align}
This describes a ``dynamical steady state'', which is characterized by a longitudinally growing cylinder of excited states of length $eEt$ and radius $\sim \sqrt{\frac{\hbar eE}{\pi v_F}}$. In contrast to the short time limit, the excitation probability decays exponentially for large momentum. This exponential decay can be explained as a tunneling effect within the WKB approach \cite{CasherPRD79}. 

Along with the analytical calculations, for comparison, we determine numerically $n_{\vp}$ and $\partial_t n_{\vp}$ by applying an explicit Runge-Kutta method \cite{ButcherNumODE} to solve the time dependent Schr\"odinger equation. In Fig.~\ref{fig:np_plane} we compare the approximations used for $n_{\vp}$ with the numerically obtained values.  

\section{Evolution of the current} \label{sec:currevol}

We are now in a position to discuss the time evolution of the current. The high energy cutoff,  $v_F \Lambda$ defines an ultrashort timescale $t_{\Lambda}=\frac{\hbar}{v_F \Lambda}$, which satisfies $t_{\Lambda}\ll t_E$ for both condensed matter\cite{BurkovPRL2011} and cold atomic\cite{HeWeyl2015} realizations of WSMs, similarly to the 2D case \cite{TarruellNat2012}. The scaling property \eqref{eq:np_scale} implies a scaling for the integrated current as
\begin{align}
\jint{c/p}{E,\Lambda}(t)=b^{-3} \jint{c/p}{b^2 E,b \Lambda}(b^{-1}t) \,.
 \end{align}
The particular choice of $b=t_E$ allows us to reveal the electric field and time dependence of the physical quantities. 
The current is expressed as $\jint{c/p}{E,\Lambda}(t)=E^{3/2} \jint{c/p}{1,t_{\Lambda}/t_E}(t/t_E)$. 
The scaling functions $j_{1,y}^{\text{c/p}}(x)$ are determined from Eqs.~(\ref{eq:jpc},\ref{eq:jpp}) after evaluating the momentum integrals with the particular form of $n_{\vp}(t)$,
\begin{align}
\jint{c}{E,\Lambda}(t)\sim E^{3/2}\begin{cases}
		    -\big(\frac{t}{t_E}\big)^3\ln{\frac{t\, t_{\Lambda}}{t_E^2}} & t\ll t_{\Lambda} \\
		    -\big(\frac{t}{t_E}\big)^3\ln{\frac{t}{t_E}} & t_{\Lambda}\ll t\ll t_E \\
		    \frac{t}{t_E}		& t_E \ll t
                  \end{cases}
\end{align}
\begin{align} \label{eq:jintp}
\jint{p}{E,\Lambda}(t)\sim  E^{3/2}\begin{cases}
		    \frac{t\, t_E}{t_{\Lambda}^2} & t\ll t_{\Lambda} \\
		    (1+\text{non-univ.})\frac{t_E}{t} 	& t_{\Lambda}\ll t\ll t_E \\
		    \text{const}		& t_E \ll t
                  \end{cases}
\end{align}
The term "non-univ." in the second line of Eq.~\eqref{eq:jintp} denotes the non-universal contribution from the high energy regularization, which dies out with increasing time, as discussed further in Eq. \eqref{eq:jpolsm}.

For $t\ll t_E$ the current is dominated by the polarization part. Because of the large weight of high energy states available to excite at ultrashort times 
$t < t_{\Lambda}$, the current is determined by the cutoff. The ultrashort time contribution of a Weyl point to the current is linear in time,
\begin{align}
 j(t)=\frac{1}{6 \pi^2}\frac{e v_F}{\hbar^3} eEt \Lambda^2 \,. 
\end{align}
This behavior has also been observed for 2D Dirac fermions  \cite{DoraPRB2010}, and can be explained using a classical picture of particles with 
effective mass $m_{i,j}^{-1}=\frac{\partial^2 \epsilon_{\vp}}{\partial p_i \partial p_j}$ accelerating in an external field satisfying Newton's equation. 
In 2D, the current saturates at $t\sim t_{\Lambda}$, and remains constant until $t\sim t_E$. In 3D the behavior is sharply different as the current 
starts to decay as $t^{-1}$ after reaching a maximal value at $t\sim t_{\Lambda}$. The precise form of the decay depends on the microscopic details 
(i.e. on the cutoff), but the exponent is a universal characteristic of Weyl physics. Imposing a sharp cutoff results in an oscillating current 
$j~\sim t^{-1}(1+\cos(t/t_{\Lambda}))$, also obtained within linear response\cite{RosensteinPRB2013}. A smooth (exponential or Gaussian) cutoff of the form $\exp(-\sqrt{2}p/\Lambda)$ or $\exp(-p^2/\Lambda^2) $ does not generate the oscillating part, and gives
\begin{align} \label{eq:jpolsm}
 j(t)=\frac{1}{6 \pi^2}\frac{e^2 E}{\hbar v_F t} F(t/t_{\Lambda}) \,. 
\end{align}
where $F(x)\sim x^2$ for $x\ll 1$ and $F(x)=1/2$ for $x\gg 1$.
The qualitative difference between the 2D and 3D cases is a consequence of their respective phase space sizes. The polarization current is a sum of contributions with oscillating signs $j\sim \int \mathrm{d}p \, \sin(2 p t) p^{d-3}$, which, by simple scaling, results in a time independent contribution in 2D, but decays as $t^{-1}$ in 3D.

\begin{figure}[h!]
\centering
\includegraphics[width=8.4cm]{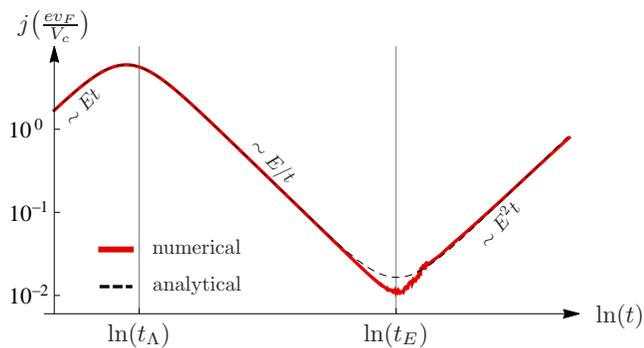}
\caption{Time evolution of the total current after switching on an electric field. The analytical curve is the sum of polarization current \eqref{eq:jpolsm}, dominant for $t\ll t_E$, and conduction current \eqref{eq:jcondLZ}, dominant for $t$.
The evolution of the number of pairs, $\kappa_1$, is shown in Fig. \ref{fig:cumulant}.}
\label{fig:current}
\end{figure}

The conduction part overtakes the polarization term at $t\sim t_E$, beyond which the current increases linearly with time and nonlinearly with electric field as
\begin{align} \label{eq:jcondLZ}
 j(t)=\frac{1}{4 \pi^3}\frac{e^3 E^2}{\hbar^2} t \,. 
\end{align}
This is simply the number of charge carriers per unit volume in the steady-state cylinder multiplied by $e v_F$. It is beyond linear response, as it depends quadratically on the external field \cite{DoraPRB2011}. 
Our analytical predictions for the current are illustrated on Fig.~\ref{fig:current}, together with the numerical results.

Bloch oscillations appear on a timescale $t_{\text{Bloch}}\sim \frac{\hbar}{e E a}$, where $a$ is the lattice constant, and our description holds for $t\ll t_{\text{Bloch}}$. 
The timescale related to the cutoff is non-universal, and both $t_E$ and $t_{\text{Bloch}}$ depend on the applied field. These three scales are in fact not independent, which can be seen in the following way. The momentum 
cutoff is proportional to the largest momentum in the system $\Lambda=\frac{1}{c} \frac{\hbar}{a}$, which relates the timescales as $t_{\Lambda}t_{\text{Bloch}}=c \,t_E^2$, where $c > 1$ is a system dependent constant describing the 
ratio of the linear size of the Brillouin zone and the validity range of Weyl physics. This also implies that in the experimentally relevant $t_{\Lambda}\ll t_E$ case, $t_E \ll t_{\text{Bloch}}$ is also satisfied, and all three regions 
appear before Bloch oscillations set in.
It is interesting to note that the maximal current is $j_{max}\sim e^2 v_FE\Lambda/\hbar^3$, which the system reaches upon leaving the classical region during the time evolution. Even in the nonlinear region in Eq. \eqref{eq:jcondLZ},
the maximal current falls to the same order of magnitude, which is in sharp contrast to 2D Dirac semimetals, where the non-linear current strongly exceeds the current from the classical region.

As the external field induces current, it also injects  energy into the system. The conduction and the polarization current decompose the total pumped energy into reversible (``work'') and irreversible (``heat''), as follows. An infinitesimal change in the energy can be written as $\mathrm{d}E=\sum_{i}(\mathrm{d}\epsilon_{i}n_{i} + \epsilon_{i} \mathrm{d}n_{i})$, where $i=(\vp,\pm)$ runs over all momenta of the two bands. The first term corresponds to the reversible work done on the system, $\mathrm{d}W=\sum_{\vp}{\partial_t \epsilon_{\vp}}(2 n_{\vp}-1) \mathrm{d}t =V E \jint{c}{}(t) \mathrm{d}t$, while the second corresponds to the heat exchange, $\mathrm{d}Q=2\sum_{\vp}\epsilon_{\vp} {\partial_t n_{\vp}}\mathrm{d}t=V E \jint{c}{}(t) \mathrm{d}t$, where we have expressed everything by the properties of the lower band. Correspondingly the work done on the system and the heat are 
\begin{align}
 W&=V E \int_0^t \mathrm{d}s \,\jint{c}{}(s)\\
 Q&=V E \int_0^t \mathrm{d}s \, \jint{p}{}(s) \,.
\end{align}
It is easy to check that the sum of the heat and work yields the total energy of the time evolved state $\Delta E=W+Q=\sum_{\vp} 2 \epsilon_{\vp} n_{\vp}$, i.e. simply the sum of the energy absorbed by the excited modes.

\section{Steady state picture from Drude theory} \label{sec:Drude}

The Drude picture provides a heuristic way to relate our results to optical conductivity studies of a WSM in the presence of impurities. In general, this is expected to work\cite{ashcroft} for (contributions to) quantities independent of the relaxation time, as e.g. the high frequency conductivity or the universal minimal conductivity of graphene\cite{castro07}. 
In this spirit, the dynamics described above holds until a characteristic time determined by an effective scattering rate $1/\tau$, and 
the zero frequency limit of the AC conductivity can be estimated by substituting time as $t\rightarrow \tau$. This results 
in the counterintuitive conclusion that in the $t_{\Lambda}<\tau<t_E$ region, the conductivity is proportional to the scattering rate, 
$\sigma(\omega\rightarrow 0) \approx \frac{e^2}{12\pi^2 \hbar v_F \tau}$, which agrees with the results of 
Ref.~\cite{AshbyPRB2014} based on Kubo formula calculations. Although this simple Drude picture works well for graphene \cite{DoraPRB2010}, 
it fails to describe the transport properties of WSMs because in 3D, the density of states at the Weyl point vanishes even in the presence of 
small amounts of disorder \cite{OminatoPRB2014}, and concomitantly the quasiparticle lifetime diverges \cite{BurkovPRL2011,HosurPRL2012}. The Drude 
picture can be rescued if we apply it to $j_{\vp}$, substituting the time variable with the momentum dependent lifetime, and then evaluating 
the integral. The scattering rate in the Boltzmann or Born approximation is $1/\tau_{\vp}=2\pi \gamma g(\epsilon_{\vp})$ \cite{BurkovPRL2011,HosurPRL2012}, 
where $g(\epsilon)=\frac{\epsilon^2}{2 \pi^2 \hbar^2 v_F^3}$ is the density of states, and $\gamma$ characterizes the scattering strength. 
In the large scattering limit $\gamma\gg \frac{\hbar v_F^2}{\Lambda}$, integrating Eq.~\eqref{eq:jpp} with $n_{\vp}(\tau_{\vp})$ from \eqref{eq:npBalazs} reproduces the results of Refs.~\cite{BurkovPRL2011,HosurPRL2012}, that is 
\begin{gather}
\sigma \sim \frac{e^2 v_F^2}{\hbar \gamma},
\end{gather}
with a different prefactor and an additional logarithmic correction $\sim\frac{e^2 v_F^2}{\hbar \gamma} \ln(\frac{\hbar v_F^2}{\gamma\Lambda})$. 
The above treatment is valid for small electric fields $eE \ll \frac{\gamma^2 \Lambda^4}{\hbar^3 v_F^3}$, when the dominant contribution to the current comes from the momenta satisfying $\tau_{\vp} \ll t_E$.  

If the scattering strength is small, such that there is enough time for the modes to go through the LZ transition, then the steady state occupation profile will be qualitatively similar to the LZ solution. As the 
quasiparticle lifetime is finite everywhere except in the close vicinity of the Weyl point, the cylinder of densely excited states will not extend to infinity, but will be characterized by a finite length 
$eE \tau_{\text{eff}}(E)$. The precise form of $\tau_{\text{eff}}$ depends on the detailed nature  of the scattering process. If there is a constant scattering rate $1/\tau$, then $\tau_{\text{eff}}=\tau$,
 but generally it will depend on the electric field. The Drude picture estimates the stationary current in the non-linear regime as 
\begin{gather}
j_{\text{stac}}=\frac{1}{4 \pi^3}\frac{e^3 E^2}{\hbar^2} \tau_{\text{eff}}(E), 
\end{gather}
and Ohm's law breaks down. The explicit $E$ dependence, however, depends strongly on the precise form of $\tau_{\text{eff}}(E)$. In case the relaxation time becomes independent of the electric field in the non-linear region,
a crossover from the $j\sim E$ linear region to a $j\sim E^2$ non-linear region is expected.

\section{Statistics of pair creation}\label{sec:statistics}

\begin{figure}[h!]
\centering
\includegraphics[width=8.4cm]{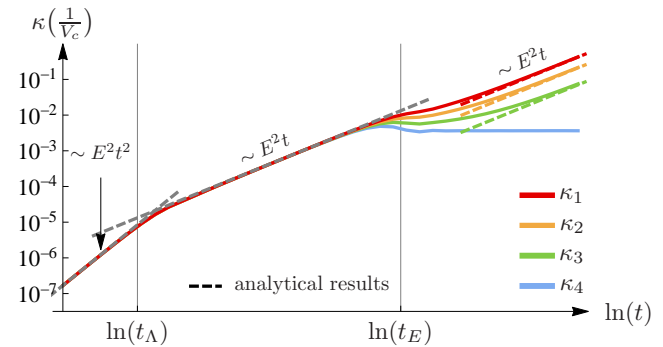}
\caption{Time evolution of the cumulants ($\kappa_{1-4}$) of electron-hole pairs per unit volume (log-log plot, numerical results). 
The cumulants coincide for $t\ll t_E$, which is a clear signature of a Poissonian distribution. The grey dashed lines show the 
$t\ll t_{\Lambda}$ and $t\gg t_{\Lambda}$ asymptotics of the analytical formula Eq.~\eqref{eq:lambda_poi}. For long times, the 
cumulants branch and follow the approximate formulae derived from Eq.~\eqref{eq:charfc_LZ} (colored dashed lines) within a time 
independent constant coming from the difference between the exact $n_{\vp}$ and the LZ approximation (Fig.~\ref{fig:np_plane}).}
\label{fig:cumulant}
\end{figure}

The expectation value and time evolution of the current is largely influenced by the number of pairs created, as follows from Eqs. \eqref{eq:jpc}, \eqref{eq:jpp}. This we now investigate in more detail. 
Although the expectation value of a quantity reveals much about underlying physical processes, fluctuations contain essential information as well and are important  to provide
 a comprehensive description of the system\cite{CampisiRMP2011}. Therefore, beyond simple expectation values, we study the fluctuations of the pairs created 
by their full distribution function. This provides a 
complementary measure to characterize the out-of-equilibrium state. 
As opposed to calculating the probability distribution function of pairs created directly, it is more 
convenient to work with the cumulant generating function (CGF) in unit volume, 
which is the logarithm of the characteristic function $\phi(\varphi)=\frac{1}{V}\ln \left\langle \exp(i \varphi \hat{N})\right\rangle$. Here, 
 $\hat{N}$ denotes the excitation number operator, and the expectation value is taken with the time evolved initial state. The CGF is expressed as sum over momentum space, 
\begin{align} \label{eq:charfc_indep}
\phi(\varphi)&=\frac{1}{V}\sum\limits_{\vp} \ln\left[1+(\exp(i \varphi)-1)n_{\vp}\right]
\end{align}
The probability density function is the inverse Fourier transform of the characteristic function, that is, $p(n)=\frac{1}{2\pi}\int \mathrm{d}\varphi\, \exp(V \phi(\varphi)-i n \varphi)$. 
For short perturbation, i.e. $t \ll t_E$ the excitations add up from an extended region in momentum space with small excitation probability. The contribution from $p \lesssim 2 e E t$, where $n_{\vp}\sim 1$, is negligible because of the small volume of the domain $\sim t^3$, and a Taylor expansion of the logarithm in Eq.~\eqref{eq:charfc_indep} gives a good approximation,
$\phi(\varphi)=(\exp(i\varphi)-1)\frac{1}{V}{\sum n_{\vp}}$.
That is, the total number of excitations per unit volume is Poissonian as $p(n)=\lambda^n\exp(-\lambda)/n!$ with 
\begin{align} \label{eq:lambda_poi}
\lambda=\frac{1}{12 \pi^2} \frac{(e E)^2 t}{\hbar^2 v_F} S_2(t/t_{\Lambda}) \,,
\end{align}
where $S_2(y)=\int_0^{y} \mathrm{d}x ~{\sin^2 x}/{x^2}=y$ for $y\ll 1$, while it saturates to ${\pi}/{2}$ for $y\gg 1$.
All cumulants of the Poisson distribution are equal to the single parameter $\lambda$. The first cumulant is the expectation value, that is, 
for $t\ll t_{\Lambda}$ the excitations are created quadratically in time, while for $t_{\Lambda}\ll t \ll t_E$, the creation rate is constant. 
This behavior is clearly seen in Fig.~\ref{fig:cumulant}, where we compare the numerically determined cumulants with the approximate solutions.

For $t_E\ll t$ the excited states are confined to a cylinder in momentum space, and substituting Eq.~\eqref{eq:npLZ} into \eqref{eq:charfc_indep} yields
\begin{align} \label{eq:charfc_LZ}
\phi(\varphi)&=- \alpha \Li_2(1-\exp(i \varphi)) \\ \label{eq:alphaLZ}
\alpha&=\frac{1}{8 \pi^3} \frac{(e E)^2 t}{\hbar^2 v_f} 
\end{align}
where $\Li_2(x)=\sum_{m=1}^{\infty}{x^m}/{m^2}$ is the dilogarithm 
function \cite{gradshteyn1980table}, in agreement with Ref.~\cite{VildanovPRB2010}. As time evolves the 
higher cumulants start to deviate from the first one, and the distribution is no longer Poissonian (see Fig.~\ref{fig:cumulant}).
 The cumulants are determined from the series expansion of the CGF, the first few being $\kappa_1=\alpha$, $\kappa_2=\alpha/2$, 
$\kappa_3=\alpha/6$, $\kappa_4=0$. Except for the variance all even cumulants vanish. There is a time independent contribution 
from the $p_x\approx 0$ and $p_x\approx eEt$ regions  in $n_{\vp}$ (see Fig.~\ref{fig:np_plane}), which is not captured in Eq.~\eqref{eq:npLZ}, which gets
overwhelmed by the time dependent terms.
Apart from this, the cumulants listed above approximate very well the numerical results (Fig.~\ref{fig:cumulant}). The peak of the distribution 
function is well captured in the central limit theorem (CLT) approximation, which states that the excitation number is Gaussian with mean $\alpha$ and 
variance $\sigma^2=\alpha/2$: $p(n)=\frac{1}{\sqrt{2\pi \alpha}}\exp\{-{(n-\alpha)^2}/{\alpha}\}$. This approximation neglects the cumulants 
higher than the second. The asymptotic decay of the distribution can be determined from the G\"{a}rtner-Ellis theorem \cite{TouchettePR2009}, which 
in this case is essentially a saddle point approximation of the inverse Fourier transform of the characteristic function. 
The probability of exciting a large number of pairs decays slower than estimated from the CLT, but still in a Gaussian manner 
$p(n)\sim \exp\{-{n^2}/{2\alpha}\}$ (note the factor 2 difference in the denominator of the exponential with respect to the Gaussian distribution).

\section{Probability of no current and the vacuum persistence
probability} \label{sec:vpp}

\begin{figure}[h!]
\centering
\includegraphics[width=8.4cm]{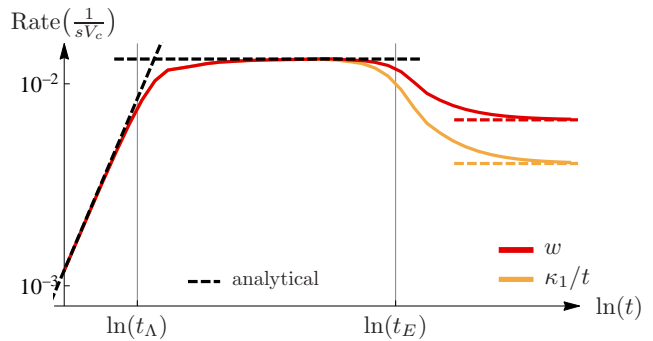}
\caption{Particle creation rate as a function of time estimated from the vacuum persistence probability and from the total number of excitations (log-log plot). The dashed lines show the results of Eqs.~(\ref{eq:lambda_poi},\ref{eq:alphaLZ}) and Eq.~\eqref{eq:w}.}
\label{fig:rate}
\end{figure}

In spite of the applied electric field, there is a finite probability of no pair creation and no induced current, also known as the vacuum persistence
probability. Analyzing the decay of this probability provides an alternative way to determine the pair-creation rate, which was used e.g. by Schwinger 
in his seminal paper \cite{SchwingerPR1951}. The vacuum persistence probability is $P_0(t)=|\left\langle \tilde{0} \right| U(t,0) \left| 0 \right\rangle |^2$, 
where $U(t,0)$ is the time evolution operator in the external field, $\left| 0 \right\rangle$ and $\left| \tilde{0} \right\rangle$ are the (Schr\"odinger) vacua 
at time $0$ and $t$ respectively. With the knowledge of $n_{\vp}$ it is expressed as 
\begin{gather}
P_0=\exp\left(-\sum_{\vp}\ln(1-n_{\vp})\right)\equiv \exp\left(-V w t\right),
\end{gather}
where  
\begin{gather} \label{eq:w}
w=\dfrac 1 t\times \left\{\begin{array}{cc}
\lambda & \textmd{ for } t \ll t_E,\\
 \dfrac{\alpha\pi^2}{6} &  \textmd{ for } t\gg t_E
\end{array}
\right.
\end{gather}
 is the rate of vacuum decay per unit volume and time, increasing as $E^2$ and being independent of time for $t\gg t_\Lambda$ and increasing linearly with time for $t\ll t_\Lambda$.
Alternatively, the pair-creation rate can also be defined as the total number of pairs created divided by the time it took, i.e. as $\kappa_1/t$. 
Nevertheless, these two definitions agree in the short time limit and only differ by a constant prefactor in the long perturbation limit (Fig.~\ref{fig:rate}). 
The vacuum persistence probability characterizes the time evolution similarly to the Loschmidt echo\cite{silva}: it measures
 the overlap of the non-equilibrium time evolved wave function $U(t,0)\left| 0 \right\rangle$ with a reference wave function, which in this case is the adiabatically evolved state.
 
So far we have assumed the initial state to be the ground state without any excitations, which describes the zero temperature response of WSMs. An arbitrary initial distribution function can be handled similarly, as long the modes with different momenta are independent, which is the case e.g. at finite temperature. Let $f(\vp)$ be the probability distribution function of having an excitation with momentum $\vp$ in the initial state. The post-quench occupation number is expressed as a weighted sum of the excitation probability of an unexcited and an excited mode as\cite{TanjiAnnPhys2009} 
$$n_{\vp}^{f}= \left[1-f(\vp-e\vE t)\right]n_{\vp} +f(\vp-e\vE t)\left[1-n_{\vp}\right] \,,$$
where $f(\vp)={1}/(\exp(\beta \epsilon_{\vp})+1)$, $\beta={1}/{k_B T}$. 
The initial number of excited states due to thermal fluctuations is $n_T\sim {1}/{(\beta \hbar v_F)^3}$, which is small near $T=0$, and does not modify qualitatively the results. This argument applies for systems with thermal initial density matrices, which are detached from the environment during time evolution. This assumption needs a thermalization time much longer than the observation time, which is usually not satisfied in condensed matter, but could be achieved with cold atoms. Similarly a small deviation in the Fermi energy from the Weyl point gives only a subleading correction.

\section{Conclusion} \label{sec:conclusion}

We have studied the nonlinear response of WSMs after switching on an external electric field before Bloch oscillations set in. 
The ultrashort time dynamics is non-universal and the current depends on the details of the band structure at high energies. The current and the number of created electron-hole pairs  grow linearly and quadratically with time, 
respectively.
 The universal properties of Weyl nodes are manifested in the intermediate and long time responses. In particular, at intermediate times, the current \emph{decays} as $1/t$ due to the interplay of the number of created pairs and
the available phase space. Particles are created at a constant rate, generating a Poissonian distribution for the statistics of the electron-hole pairs. 
At long times, the particle creation rate takes on a constant value again, but the current starts to increase again in time because of the increasingly large number of pairs moving in the same direction. 
The distribution function of excitations crosses over from a Poissonian profile to a Gaussian distribution, which follows from the central limit theorem, applicable in the long time limit due to the large number of pairs created.
The real time evolution of the current is translated to the conductivity of disordered WSMs within a 
generalized Drude picture, reproducing the results of previous calculations with different methods.
This is a remarkable example of a problem from high energy physics which
naturally corresponds to one in condensed matter physics with a separate
set of observables, and which allows an exquisitely detailed analysis,
thus holding the promise of a detailed experimental study in a tabletop
experiment.

\begin{acknowledgments}
This research has been  supported by the Hungarian Scientific  Research Funds Nos. K101244, K105149, K108676, by the 
ERC Grant Nr. ERC-259374-Sylo and by the Bolyai Program of the Hungarian Academy of Sciences.
\end{acknowledgments}

\bibliographystyle{apsrev}

\bibliography{refgraph}

\begin{thebibliography}{33}
\expandafter\ifx\csname natexlab\endcsname\relax\def\natexlab#1{#1}\fi
\expandafter\ifx\csname bibnamefont\endcsname\relax
  \def\bibnamefont#1{#1}\fi
\expandafter\ifx\csname bibfnamefont\endcsname\relax
  \def\bibfnamefont#1{#1}\fi
\expandafter\ifx\csname citenamefont\endcsname\relax
  \def\citenamefont#1{#1}\fi
\expandafter\ifx\csname url\endcsname\relax
  \def\url#1{\texttt{#1}}\fi
\expandafter\ifx\csname urlprefix\endcsname\relax\def\urlprefix{URL }\fi
\providecommand{\bibinfo}[2]{#2}
\providecommand{\eprint}[2][]{\url{#2}}

\bibitem[{\citenamefont{Wan et~al.}(2011)\citenamefont{Wan, Turner, Vishwanath,
  and Savrasov}}]{WanPRB2011}
\bibinfo{author}{\bibfnamefont{X.}~\bibnamefont{Wan}},
  \bibinfo{author}{\bibfnamefont{A.~M.} \bibnamefont{Turner}},
  \bibinfo{author}{\bibfnamefont{A.}~\bibnamefont{Vishwanath}},
  \bibnamefont{and} \bibinfo{author}{\bibfnamefont{S.~Y.}
  \bibnamefont{Savrasov}}, \bibinfo{journal}{Phys. Rev. B}
  \textbf{\bibinfo{volume}{83}}, \bibinfo{pages}{205101}
  (\bibinfo{year}{2011}).

\bibitem[{\citenamefont{Burkov and Balents}(2011)}]{BurkovPRL2011}
\bibinfo{author}{\bibfnamefont{A.~A.} \bibnamefont{Burkov}} \bibnamefont{and}
  \bibinfo{author}{\bibfnamefont{L.}~\bibnamefont{Balents}},
  \bibinfo{journal}{Phys. Rev. Lett.} \textbf{\bibinfo{volume}{107}},
  \bibinfo{pages}{127205} (\bibinfo{year}{2011}).

\bibitem[{\citenamefont{Hosur et~al.}(2012)\citenamefont{Hosur, Parameswaran,
  and Vishwanath}}]{HosurPRL2012}
\bibinfo{author}{\bibfnamefont{P.}~\bibnamefont{Hosur}},
  \bibinfo{author}{\bibfnamefont{S.~A.} \bibnamefont{Parameswaran}},
  \bibnamefont{and}
  \bibinfo{author}{\bibfnamefont{A.}~\bibnamefont{Vishwanath}},
  \bibinfo{journal}{Phys. Rev. Lett.} \textbf{\bibinfo{volume}{108}},
  \bibinfo{pages}{046602} (\bibinfo{year}{2012}).

\bibitem[{\citenamefont{{Burkov}}(2015)}]{BurkovJPCM2015}
\bibinfo{author}{\bibfnamefont{A.~A.} \bibnamefont{{Burkov}}},
  \bibinfo{journal}{Journal of Physics: Condensed Matter}
  \textbf{\bibinfo{volume}{27}}, \bibinfo{pages}{113201}
  (\bibinfo{year}{2015}).

\bibitem[{\citenamefont{Schwinger}(1951)}]{SchwingerPR1951}
\bibinfo{author}{\bibfnamefont{J.}~\bibnamefont{Schwinger}},
  \bibinfo{journal}{Phys. Rev.} \textbf{\bibinfo{volume}{82}},
  \bibinfo{pages}{664} (\bibinfo{year}{1951}).

\bibitem[{\citenamefont{Polkovnikov}(2005)}]{PolkovnikovPRB2005}
\bibinfo{author}{\bibfnamefont{A.}~\bibnamefont{Polkovnikov}},
  \bibinfo{journal}{Phys. Rev. B} \textbf{\bibinfo{volume}{72}},
  \bibinfo{pages}{161201} (\bibinfo{year}{2005}).

\bibitem[{\citenamefont{Zurek et~al.}(2005)\citenamefont{Zurek, Dorner, and
  Zoller}}]{ZurekPRL2005}
\bibinfo{author}{\bibfnamefont{W.~H.} \bibnamefont{Zurek}},
  \bibinfo{author}{\bibfnamefont{U.}~\bibnamefont{Dorner}}, \bibnamefont{and}
  \bibinfo{author}{\bibfnamefont{P.}~\bibnamefont{Zoller}},
  \bibinfo{journal}{Phys. Rev. Lett.} \textbf{\bibinfo{volume}{95}},
  \bibinfo{pages}{105701} (\bibinfo{year}{2005}).

\bibitem[{\citenamefont{Damski and Zurek}(2006)}]{DamskiPRA2006}
\bibinfo{author}{\bibfnamefont{B.}~\bibnamefont{Damski}} \bibnamefont{and}
  \bibinfo{author}{\bibfnamefont{W.~H.} \bibnamefont{Zurek}},
  \bibinfo{journal}{Phys. Rev. A} \textbf{\bibinfo{volume}{73}},
  \bibinfo{pages}{063405} (\bibinfo{year}{2006}).

\bibitem[{\citenamefont{{Tarruell} et~al.}(2012)\citenamefont{{Tarruell},
  {Greif}, {Uehlinger}, {Jotzu}, and {Esslinger}}}]{TarruellNat2012}
\bibinfo{author}{\bibfnamefont{L.}~\bibnamefont{{Tarruell}}},
  \bibinfo{author}{\bibfnamefont{D.}~\bibnamefont{{Greif}}},
  \bibinfo{author}{\bibfnamefont{T.}~\bibnamefont{{Uehlinger}}},
  \bibinfo{author}{\bibfnamefont{G.}~\bibnamefont{{Jotzu}}}, \bibnamefont{and}
  \bibinfo{author}{\bibfnamefont{T.}~\bibnamefont{{Esslinger}}},
  \bibinfo{journal}{\nat} \textbf{\bibinfo{volume}{483}}, \bibinfo{pages}{302}
  (\bibinfo{year}{2012}).

\bibitem[{\citenamefont{{He} et~al.}(2015)\citenamefont{{He}, {Zhang}, and
  {Law}}}]{HeWeyl2015}
\bibinfo{author}{\bibfnamefont{W.-Y.} \bibnamefont{{He}}},
  \bibinfo{author}{\bibfnamefont{S.}~\bibnamefont{{Zhang}}}, \bibnamefont{and}
  \bibinfo{author}{\bibfnamefont{K.~T.} \bibnamefont{{Law}}},
  \bibinfo{journal}{ArXiv e-prints}  (\bibinfo{year}{2015}),
  \eprint{1501.02348}.

\bibitem[{\citenamefont{Nazarov}(2003)}]{nazarov2003quantum}
\bibinfo{author}{\bibfnamefont{Y.~V.} \bibnamefont{Nazarov}},
  \emph{\bibinfo{title}{Quantum Noise in Mesoscopic Physics}}
  (\bibinfo{publisher}{Springer}, \bibinfo{year}{2003}).

\bibitem[{\citenamefont{Esposito et~al.}(2009)\citenamefont{Esposito, Harbola,
  and Mukamel}}]{EspositoRMP2009}
\bibinfo{author}{\bibfnamefont{M.}~\bibnamefont{Esposito}},
  \bibinfo{author}{\bibfnamefont{U.}~\bibnamefont{Harbola}}, \bibnamefont{and}
  \bibinfo{author}{\bibfnamefont{S.}~\bibnamefont{Mukamel}},
  \bibinfo{journal}{Rev. Mod. Phys.} \textbf{\bibinfo{volume}{81}},
  \bibinfo{pages}{1665} (\bibinfo{year}{2009}).

\bibitem[{\citenamefont{Maisi et~al.}(2014)\citenamefont{Maisi, Kambly, Flindt,
  and Pekola}}]{MaisiPRL2014}
\bibinfo{author}{\bibfnamefont{V.~F.} \bibnamefont{Maisi}},
  \bibinfo{author}{\bibfnamefont{D.}~\bibnamefont{Kambly}},
  \bibinfo{author}{\bibfnamefont{C.}~\bibnamefont{Flindt}}, \bibnamefont{and}
  \bibinfo{author}{\bibfnamefont{J.~P.} \bibnamefont{Pekola}},
  \bibinfo{journal}{Phys. Rev. Lett.} \textbf{\bibinfo{volume}{112}},
  \bibinfo{pages}{036801} (\bibinfo{year}{2014}).

\bibitem[{\citenamefont{Malossi et~al.}(2014)\citenamefont{Malossi, Valado,
  Scotto, Huillery, Pillet, Ciampini, Arimondo, and Morsch}}]{MalossiPRL2014}
\bibinfo{author}{\bibfnamefont{N.}~\bibnamefont{Malossi}},
  \bibinfo{author}{\bibfnamefont{M.~M.} \bibnamefont{Valado}},
  \bibinfo{author}{\bibfnamefont{S.}~\bibnamefont{Scotto}},
  \bibinfo{author}{\bibfnamefont{P.}~\bibnamefont{Huillery}},
  \bibinfo{author}{\bibfnamefont{P.}~\bibnamefont{Pillet}},
  \bibinfo{author}{\bibfnamefont{D.}~\bibnamefont{Ciampini}},
  \bibinfo{author}{\bibfnamefont{E.}~\bibnamefont{Arimondo}}, \bibnamefont{and}
  \bibinfo{author}{\bibfnamefont{O.}~\bibnamefont{Morsch}},
  \bibinfo{journal}{Phys. Rev. Lett.} \textbf{\bibinfo{volume}{113}},
  \bibinfo{pages}{023006} (\bibinfo{year}{2014}).

\bibitem[{\citenamefont{Flindt et~al.}(2009)\citenamefont{Flindt, Fricke,
  Hohls, Novotny, Netocny, Brandes, and Haug}}]{FlindtPNAS2009}
\bibinfo{author}{\bibfnamefont{C.}~\bibnamefont{Flindt}},
  \bibinfo{author}{\bibfnamefont{C.}~\bibnamefont{Fricke}},
  \bibinfo{author}{\bibfnamefont{F.}~\bibnamefont{Hohls}},
  \bibinfo{author}{\bibfnamefont{T.}~\bibnamefont{Novotny}},
  \bibinfo{author}{\bibfnamefont{K.}~\bibnamefont{Netocny}},
  \bibinfo{author}{\bibfnamefont{T.}~\bibnamefont{Brandes}}, \bibnamefont{and}
  \bibinfo{author}{\bibfnamefont{R.~J.} \bibnamefont{Haug}},
  \bibinfo{journal}{Proceedings of the National Academy of Sciences}
  \textbf{\bibinfo{volume}{106}}, \bibinfo{pages}{10116}
  (\bibinfo{year}{2009}).

\bibitem[{\citenamefont{Gavrilov and Gitman}(1996)}]{GavrilovPRD96}
\bibinfo{author}{\bibfnamefont{S.~P.} \bibnamefont{Gavrilov}} \bibnamefont{and}
  \bibinfo{author}{\bibfnamefont{D.~M.} \bibnamefont{Gitman}},
  \bibinfo{journal}{Phys. Rev. D} \textbf{\bibinfo{volume}{53}},
  \bibinfo{pages}{7162} (\bibinfo{year}{1996}).

\bibitem[{\citenamefont{Vitanov and Garraway}(1996)}]{VitanovPRA96}
\bibinfo{author}{\bibfnamefont{N.~V.} \bibnamefont{Vitanov}} \bibnamefont{and}
  \bibinfo{author}{\bibfnamefont{B.~M.} \bibnamefont{Garraway}},
  \bibinfo{journal}{Phys. Rev. A} \textbf{\bibinfo{volume}{53}},
  \bibinfo{pages}{4288} (\bibinfo{year}{1996}).

\bibitem[{\citenamefont{Tanji}(2009)}]{TanjiAnnPhys2009}
\bibinfo{author}{\bibfnamefont{N.}~\bibnamefont{Tanji}},
  \bibinfo{journal}{Annals of Physics} \textbf{\bibinfo{volume}{324}},
  \bibinfo{pages}{1691 } (\bibinfo{year}{2009}), ISSN
  \bibinfo{issn}{0003-4916}.

\bibitem[{\citenamefont{D\'ora and Moessner}(2010)}]{DoraPRB2010}
\bibinfo{author}{\bibfnamefont{B.}~\bibnamefont{D\'ora}} \bibnamefont{and}
  \bibinfo{author}{\bibfnamefont{R.}~\bibnamefont{Moessner}},
  \bibinfo{journal}{Phys. Rev. B} \textbf{\bibinfo{volume}{81}},
  \bibinfo{pages}{165431} (\bibinfo{year}{2010}).

\bibitem[{\citenamefont{D\'ora and Moessner}(2011)}]{DoraPRB2011}
\bibinfo{author}{\bibfnamefont{B.}~\bibnamefont{D\'ora}} \bibnamefont{and}
  \bibinfo{author}{\bibfnamefont{R.}~\bibnamefont{Moessner}},
  \bibinfo{journal}{Phys. Rev. B} \textbf{\bibinfo{volume}{83}},
  \bibinfo{pages}{073403} (\bibinfo{year}{2011}).

\bibitem[{\citenamefont{Ashcroft and Mermin}(1976)}]{ashcroft}
\bibinfo{author}{\bibfnamefont{N.~W.} \bibnamefont{Ashcroft}} \bibnamefont{and}
  \bibinfo{author}{\bibfnamefont{N.~D.} \bibnamefont{Mermin}},
  \emph{\bibinfo{title}{Solid State Physics}} (\bibinfo{publisher}{Saunders
  College Publishing}, \bibinfo{address}{Philadelphia}, \bibinfo{year}{1976}).

\bibitem[{\citenamefont{Hallin and Liljenberg}(1995)}]{HallinPRD1995}
\bibinfo{author}{\bibfnamefont{J.}~\bibnamefont{Hallin}} \bibnamefont{and}
  \bibinfo{author}{\bibfnamefont{P.}~\bibnamefont{Liljenberg}},
  \bibinfo{journal}{Phys. Rev. D} \textbf{\bibinfo{volume}{52}},
  \bibinfo{pages}{1150} (\bibinfo{year}{1995}).

\bibitem[{\citenamefont{Casher et~al.}(1979)\citenamefont{Casher, Neuberger,
  and Nussinov}}]{CasherPRD79}
\bibinfo{author}{\bibfnamefont{A.}~\bibnamefont{Casher}},
  \bibinfo{author}{\bibfnamefont{H.}~\bibnamefont{Neuberger}},
  \bibnamefont{and} \bibinfo{author}{\bibfnamefont{S.}~\bibnamefont{Nussinov}},
  \bibinfo{journal}{Phys. Rev. D} \textbf{\bibinfo{volume}{20}},
  \bibinfo{pages}{179} (\bibinfo{year}{1979}).

\bibitem[{\citenamefont{Butcher}(2008)}]{ButcherNumODE}
\bibinfo{author}{\bibfnamefont{J.~C.} \bibnamefont{Butcher}},
  \emph{\bibinfo{title}{Numerical Methods for Ordinary Differential Equations}}
  (\bibinfo{publisher}{John Wiley \& Sons Ltd.}, \bibinfo{year}{2008}),
  \bibinfo{edition}{2nd} ed.

\bibitem[{\citenamefont{Rosenstein and Lewkowicz}(2013)}]{RosensteinPRB2013}
\bibinfo{author}{\bibfnamefont{B.}~\bibnamefont{Rosenstein}} \bibnamefont{and}
  \bibinfo{author}{\bibfnamefont{M.}~\bibnamefont{Lewkowicz}},
  \bibinfo{journal}{Phys. Rev. B} \textbf{\bibinfo{volume}{88}},
  \bibinfo{pages}{045108} (\bibinfo{year}{2013}).

\bibitem[{\citenamefont{{Castro Neto} et~al.}(2009)\citenamefont{{Castro Neto},
  Guinea, Peres, Novoselov, and Geim}}]{castro07}
\bibinfo{author}{\bibfnamefont{A.~H.} \bibnamefont{{Castro Neto}}},
  \bibinfo{author}{\bibfnamefont{F.}~\bibnamefont{Guinea}},
  \bibinfo{author}{\bibfnamefont{N.~M.~R.} \bibnamefont{Peres}},
  \bibinfo{author}{\bibfnamefont{K.~S.} \bibnamefont{Novoselov}},
  \bibnamefont{and} \bibinfo{author}{\bibfnamefont{A.~K.} \bibnamefont{Geim}},
  \bibinfo{journal}{Rev. Mod. Phys.} \textbf{\bibinfo{volume}{81}},
  \bibinfo{pages}{109} (\bibinfo{year}{2009}).

\bibitem[{\citenamefont{Ashby and Carbotte}(2014)}]{AshbyPRB2014}
\bibinfo{author}{\bibfnamefont{P.~E.~C.} \bibnamefont{Ashby}} \bibnamefont{and}
  \bibinfo{author}{\bibfnamefont{J.~P.} \bibnamefont{Carbotte}},
  \bibinfo{journal}{Phys. Rev. B} \textbf{\bibinfo{volume}{89}},
  \bibinfo{pages}{245121} (\bibinfo{year}{2014}).

\bibitem[{\citenamefont{Ominato and Koshino}(2014)}]{OminatoPRB2014}
\bibinfo{author}{\bibfnamefont{Y.}~\bibnamefont{Ominato}} \bibnamefont{and}
  \bibinfo{author}{\bibfnamefont{M.}~\bibnamefont{Koshino}},
  \bibinfo{journal}{Phys. Rev. B} \textbf{\bibinfo{volume}{89}},
  \bibinfo{pages}{054202} (\bibinfo{year}{2014}).

\bibitem[{\citenamefont{Campisi et~al.}(2011)\citenamefont{Campisi, H\"anggi,
  and Talkner}}]{CampisiRMP2011}
\bibinfo{author}{\bibfnamefont{M.}~\bibnamefont{Campisi}},
  \bibinfo{author}{\bibfnamefont{P.}~\bibnamefont{H\"anggi}}, \bibnamefont{and}
  \bibinfo{author}{\bibfnamefont{P.}~\bibnamefont{Talkner}},
  \bibinfo{journal}{Rev. Mod. Phys.} \textbf{\bibinfo{volume}{83}},
  \bibinfo{pages}{771} (\bibinfo{year}{2011}).

\bibitem[{\citenamefont{Gradshteyn and Ryzhik}(2014)}]{gradshteyn1980table}
\bibinfo{author}{\bibfnamefont{I.~S.} \bibnamefont{Gradshteyn}}
  \bibnamefont{and} \bibinfo{author}{\bibfnamefont{I.}~\bibnamefont{Ryzhik}},
  \emph{\bibinfo{title}{Table of Integrals, Series, and Products}}
  (\bibinfo{publisher}{Academic Press}, \bibinfo{year}{2014}),
  \bibinfo{edition}{8th} ed.

\bibitem[{\citenamefont{Vildanov}(2010)}]{VildanovPRB2010}
\bibinfo{author}{\bibfnamefont{N.~M.} \bibnamefont{Vildanov}},
  \bibinfo{journal}{Phys. Rev. B} \textbf{\bibinfo{volume}{82}},
  \bibinfo{pages}{033101} (\bibinfo{year}{2010}).

\bibitem[{\citenamefont{Touchette}(2009)}]{TouchettePR2009}
\bibinfo{author}{\bibfnamefont{H.}~\bibnamefont{Touchette}},
  \bibinfo{journal}{Physics Reports} \textbf{\bibinfo{volume}{478}},
  \bibinfo{pages}{1 } (\bibinfo{year}{2009}), ISSN \bibinfo{issn}{0370-1573}.

\bibitem[{\citenamefont{Silva}(2008)}]{silva}
\bibinfo{author}{\bibfnamefont{A.}~\bibnamefont{Silva}},
  \bibinfo{journal}{Phys. Rev. Lett.} \textbf{\bibinfo{volume}{101}},
  \bibinfo{pages}{120603} (\bibinfo{year}{2008}).

\end{thebibliography}

\end{document}